\documentclass[a4paper,11pt]{article}
\usepackage{pos}
\usepackage{bm}
\usepackage{booktabs}
\usepackage{tikz}
\usepackage{physics}
\usetikzlibrary{patterns}
\usepackage{pgfplots}
\usepackage{float}
\usepackage{subcaption}
\usepackage[utf8]{inputenc}
\DeclareUnicodeCharacter{2212}{-}
\DeclareUnicodeCharacter{22C6}{*}
\DeclareUnicodeCharacter{00A0}{α}

\title{Weak and Higgs physics from the lattice}
\ShortTitle{}

\author*[a]{Sofie Martins}
\author[a]{Patrick Jenny}
\author[a]{Axel Maas}
\author[a]{Georg Wieland}

\affiliation[a]{Institute of Physics, NAWI Graz, University of Graz,
Universitätsplatz 5, 8010 Graz, Austria}

\emailAdd{sofie.martins@uni-graz.at}

\abstract{The manifestly gauge-invariant and non-perturbatively complete lattice formulation of the weak interactions and the Brout-Englert-Higgs effect is connected to the usual perturbative description in phenomenology via the Fr\"ohlich-Morchio-Strocchi mechanism. However, slight differences between the two have been observed, which can potentially be accounted for by augmenting perturbation theory. We report on our ongoing lattice investigations of these additional effects using a setup with two generations of leptons coupled vectorially to the gauge-Higgs system. We explore the spectrum, inner structure in terms of weak (quasi-)PDFs, and spectral functions of the system to eventually compare cross sections to experimental results.}

\FullConference{%
The 42nd International Symposium on Lattice Field Theory,\\
2nd-8th November, 2025,\\
Tata Institute of Fundamental Research, Mumbai, India}

\begin{document}
\maketitle
	
\section{Introduction}

Electroweak phenomenology using perturbation theory around the Higgs vacuum expectation value is a fantastically successful program. However, at a more conceptual level, the notion of electroweak symmetry breaking is subtle. Gauge symmetries cannot break spontaneously, by virtue of Elitzur's theorem, while this notion is at the heart of the expansion point of perturbation theory. Any lattice formulation, being manifestly gauge-invariant, has thus to contend with the fact that quantities, like the Higgs vacuum expectation value, will vanish in the absence of gauge fixing. This apparent conundrum was resolved by Fr\"ohlich, Morchio, and Strocchi (FMS) \cite{Frohlich:1980gj, Frohlich:1981yi}. They showed that the issue lies solely in the definition of the asymptotic states, which must be manifestly gauge-invariant and thus necessarily encoded by local composite operators, akin to the hadron operators of QCD. Matrix elements, and thus transition amplitudes making up cross sections, need to be formulated in terms of such local composite operators.

After gauge fixing, the Brout-Englert-Higgs effect allows us to rewrite the matrix elements as a finite sum ordered by powers of the Higgs vacuum expectation value. In the Standard Model, the term with the highest power of the Higgs vacuum expectation value is standard perturbation theory, while the remaining terms augment the results to establish gauge invariance. In particular, the spectrum of the Standard Model is uniquely mapped one-to-one to composite states, yielding identical masses and widths. This is called the FMS mechanism. This mechanism shows conceptually why standard perturbation theory is capable of describing experimental results so well, and moreover paves the way for creating an augmented perturbation theory (APT), covering the whole expression. This is reviewed in detail in \cite{Maas:2017wzi}, with explicit calculations at NLO in \cite{Dudal:2020uwb,Maas:2020kda}.

The remaining terms are only kinematically suppressed and can thus potentially alter predictions  \cite{Maas:2017wzi,Dudal:2020uwb,Maas:2020kda,Maas:2022gdb}. It appears possible that the differences can be accounted for by the augmentation of perturbation theory \cite{Maas:2017wzi,Dudal:2020uwb,Maas:2020kda,Maas:2024hcn}, but this is not obvious. Thus, a systematic, non-perturbative treatment is called for, and, if non-perturbative contributions should exist, even required. This is the purpose of the present investigation. The novel ingredient, after early explorations \cite{Shrock:1986fg}, is that two generations of dynamical fermions are included, continuing previous lattice investigations of the pure bosonic sector (see \cite{Maas:2017wzi,Maas:2023emb} for an overview) and quenched fermions \cite{Afferrante:2020fhd}. Of course, on the lattice at the moment our fermions will only be vectorially coupled, and the mass spectrum needs to be compressed; thus, our investigations are aimed at a qualitative understanding of the mechanisms and scales. A detailed discussion of the background of our proxy theory with one generation is presented in \cite{Afferrante:2020fhd}.

Our investigations are multipronged. As a foundational investigation, we investigate the spectrum in section \ref{s:spectrum}. The structure of states is addressed by investigating (quasi-)PDFs in section \ref{s:pdf}. Finally, to move towards qualitative and order-of-magnitude estimates for experiments, we consider cross sections in section \ref{s:xs}. This program is ultimately necessary, as otherwise a discovery of the predicted effects could be easily misinterpreted as signals of physics beyond the Standard Model. In addition, similar effects occur in many scenarios beyond the Standard Model \cite{Maas:2017wzi,Maas:2023emb}, and thus establishing this framework would be of utmost importance for model building.

\begin{figure}[H]
\centering
\begin{minipage}[c]{0.25\linewidth}
\flushright
\includegraphics[width=\linewidth]{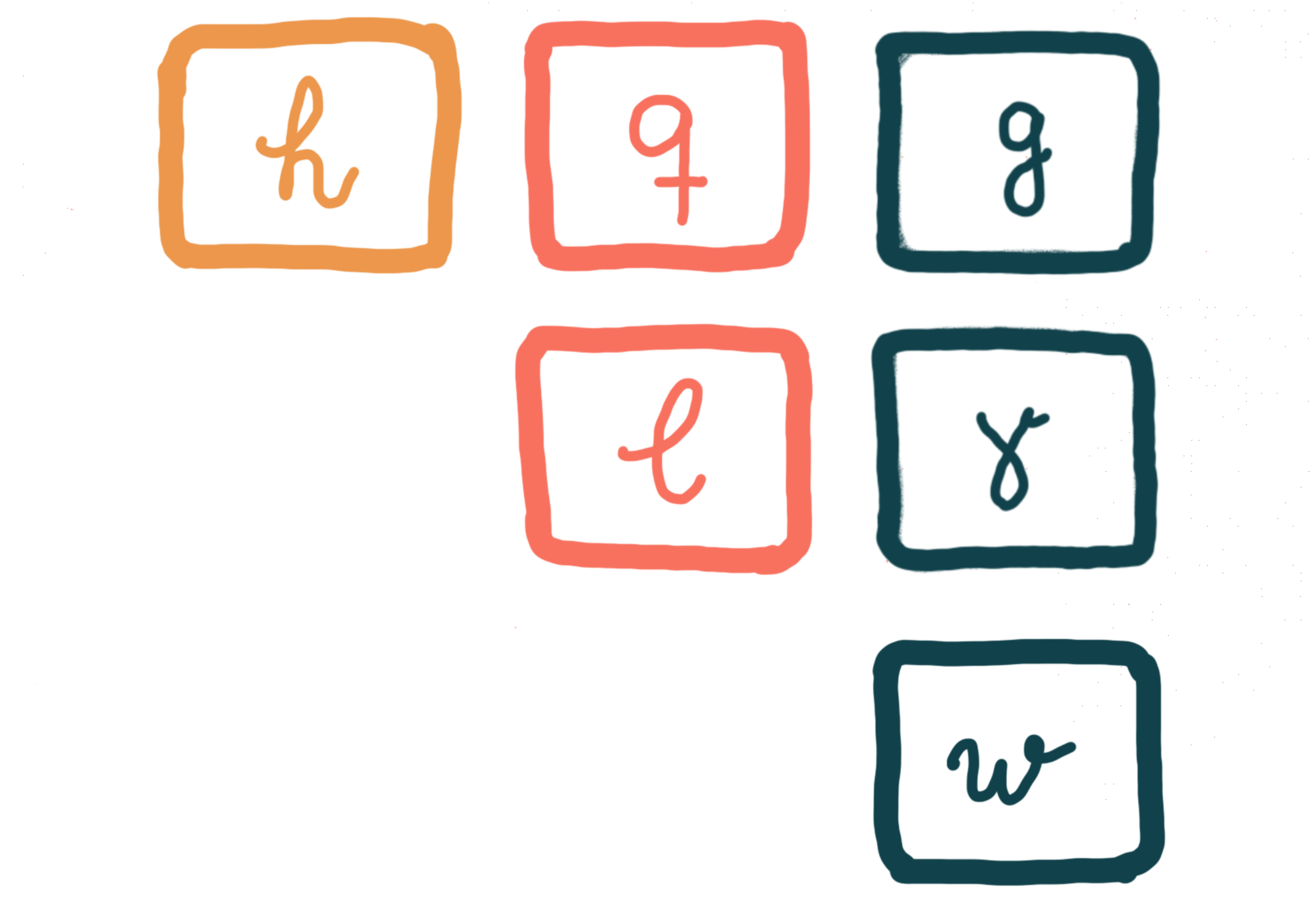}
\end{minipage}%
$\rightarrow$
\begin{minipage}[c]{0.5\linewidth}
\flushleft
\includegraphics[width=\linewidth]{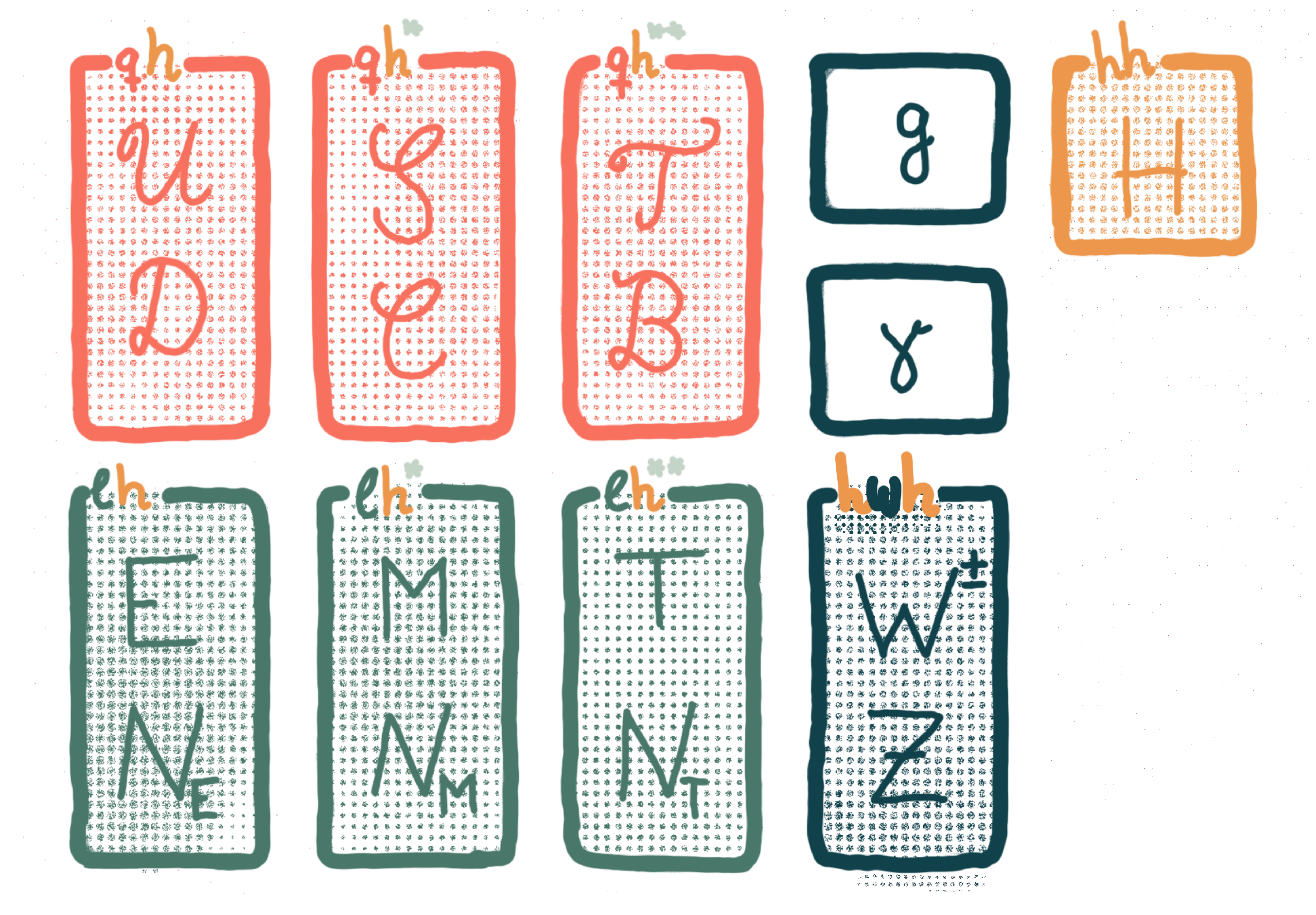}
\end{minipage}
\caption{The Standard Model reproduced from FMS. The fundamental particles (denoted by lowercase letters) are replaced by physical, gauge-invariant states. The shaded particles then appear as singlets, doublets, or triplets of the residual global \textit{custodial} symmetry \cite{Maas:2017wzi}.}
\label{fig:excited-state-SM}
\end{figure}

While a manifestly gauge-invariant treatment of the Standard Model can be accomplished with APT by using a fundamental field for each Standard Model particle, there is an even more drastic possibility. It is consistent with all available data so far that the quark and lepton generations beyond the first are actually excited states of their now-composite first-generation versions \cite{Egger:2017tkd}. This is illustrated in Fig.~\ref{fig:excited-state-SM}. Such excited states are necessarily of a non-perturbative nature. But this could not only explain the Standard Model mass hierarchies but even allow us to calculate the masses of the second- and third-generation particles, along with the CKM and PMNS matrix elements. The extension of the spectral investigation in section \ref{s:spectrum}, in particular using a variational analysis and spectral methods, will allow us to test the idea, at least for plausibility. In this context, it is noteworthy that in a similar theory hints for such an excitation spectrum have been claimed \cite{Greensite:2020lmh}.

We report here preliminary and exploratory results on this path. More systematic results will become available in the near future. However, the present ones already provide a first proof of principle for such an investigation.

\section{Simulation Setup}

The simulations are conducted using \texttt{HiRep} \cite{DelDebbio:2008zf, Drach:2025eyg}, using HMC. The action is composed of an $SU(2)$ Wilson gauge action, a complex doublet scalar field \cite{Hansen:2017mrt}, and two unimproved, degenerate Wilson fermions coupled vectorially. The latter play the role of the gauged left-handed 1st-generation (electron and electron neutrino) and 2nd-generation (muon and muon neutrino) doublet in the Standard Model \cite{Afferrante:2020fhd}. Since the electron and muon doublet have degenerate masses, this imposes an exact symmetry between the lepton generations. There is also a global $SU(2)$ symmetry associated with the Higgs doublet, which we call custodial symmetry. We expect that the concession of heavy, degenerate fermions towards computational affordability will not affect our ability to qualitatively analyze the excited-state spectrum and scattering cross sections, up to combinatorial factors and 'heavy-flavor' corrections. An overview of ensembles used for the results in this proceedings can be found in Table~\ref{tbl:ensembles}. We note that, in this theory, systematic effects are usually smaller than in QCD-like theories \cite{Maas:2014pba,Afferrante:2020fhd}.

\begin{table}[H]
\centering 
\begin{tabular}{lrrrrr}
\toprule
Label & $L$ &$\beta$ & $\kappa_\mathrm{s}$ & $\lambda_\mathrm{s}$ & $\kappa_\mathrm{F}$ \\
\midrule
Quenched & 16 & 2.7894 & 0.2984 & 1.317 & -  \\
Borderline   & 12, 20     & 2.8859 & 0.2981 & 1.334  & 0.111-0.1333  \\
Resonance    & 12, 16     & 4.0    & 0.3    & 1.0    & 0.0781-0.1220  \\
Stable Higgs & 12, 28, 32 & 8.0    & 0.131  & 0.0033 & 0.1160  \\
QCD-like & 12 & 2.3095 & 0.2668 & 0.5254 & 0.1000-0.1500 \\
\bottomrule
\end{tabular}
\caption{Overview of parameters for ensembles used for the results in sections \ref{s:spectrum}-\ref{s:xs}. The lattice dimensions are $L^4$ with $L$ listed above.}
\label{tbl:ensembles}
\end{table}

\section{Mass hierarchies and phase structure}\label{s:spectrum}

We consider the following operators 
\begin{equation}
O_{\mathrm{Higgs}} = \mathrm{tr}(X^{\dagger}X),\qquad
O_{W} = \mathrm{tr}(\tau^{a}X^{\dagger}D_{\mu}X),\qquad
\Psi = X^{\dagger}\psi~\mathrm{with}~\psi = (\nu_l, l),\label{ops}
\end{equation}
for the extraction of the spectrum, see \cite{Afferrante:2020fhd,Wieland:2025tog}. They correspond to a singlet scalar (the physical Higgs), a vector custodial triplet (the physical W/Z), and a generation and custodial fermionic doublet (the physical leptons), see Fig.~\ref{fig:excited-state-SM}.

\begin{figure}
\centering
\begin{minipage}[t]{0.45\linewidth}
\centering
\includegraphics[width=\linewidth]{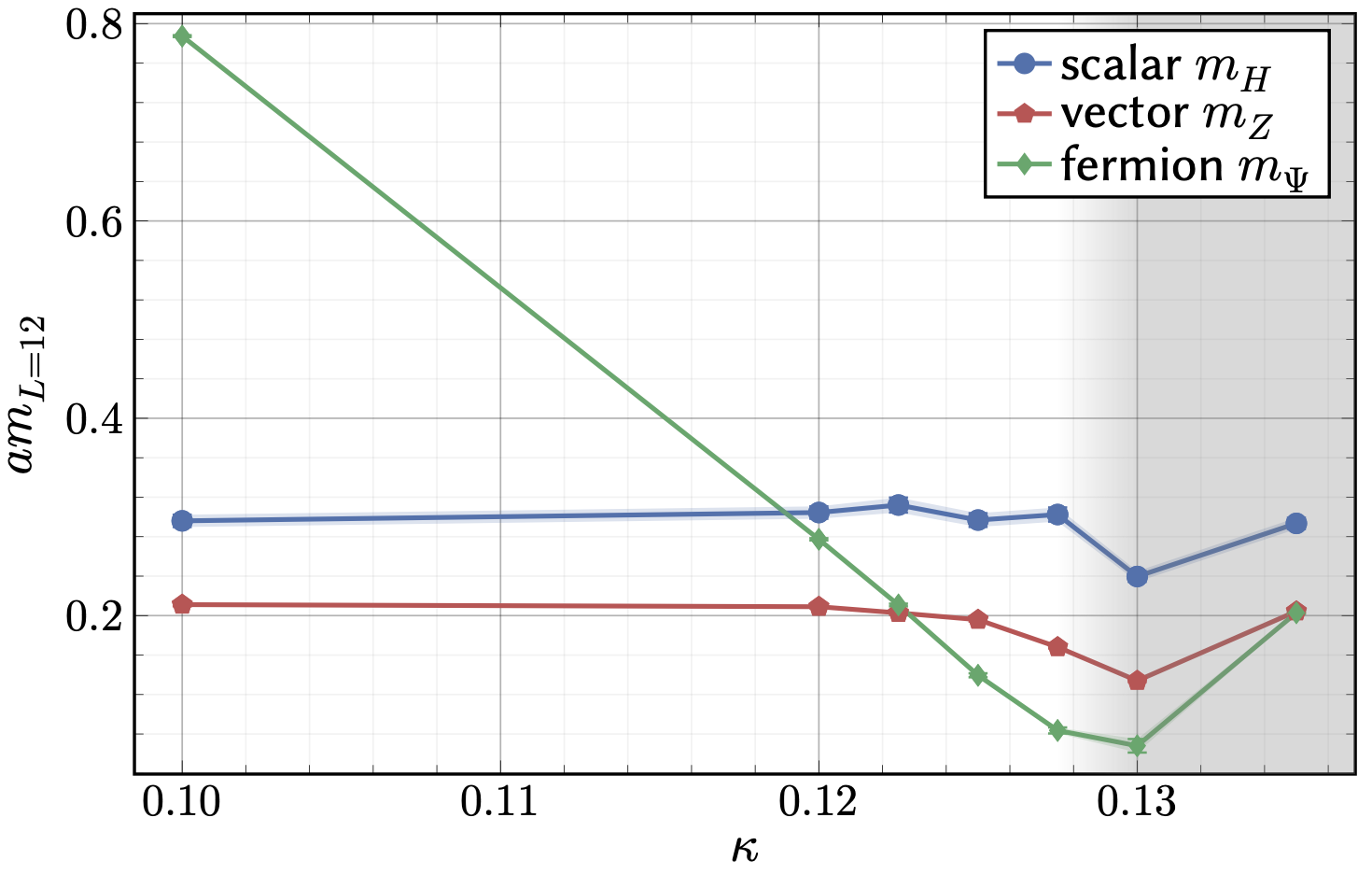}
\subcaption{Stable Higgs mass hierarchy}
\label{fig:stablehiggs}
\end{minipage}%
\begin{minipage}[t]{0.45\linewidth}
\centering
\includegraphics[width=\linewidth]{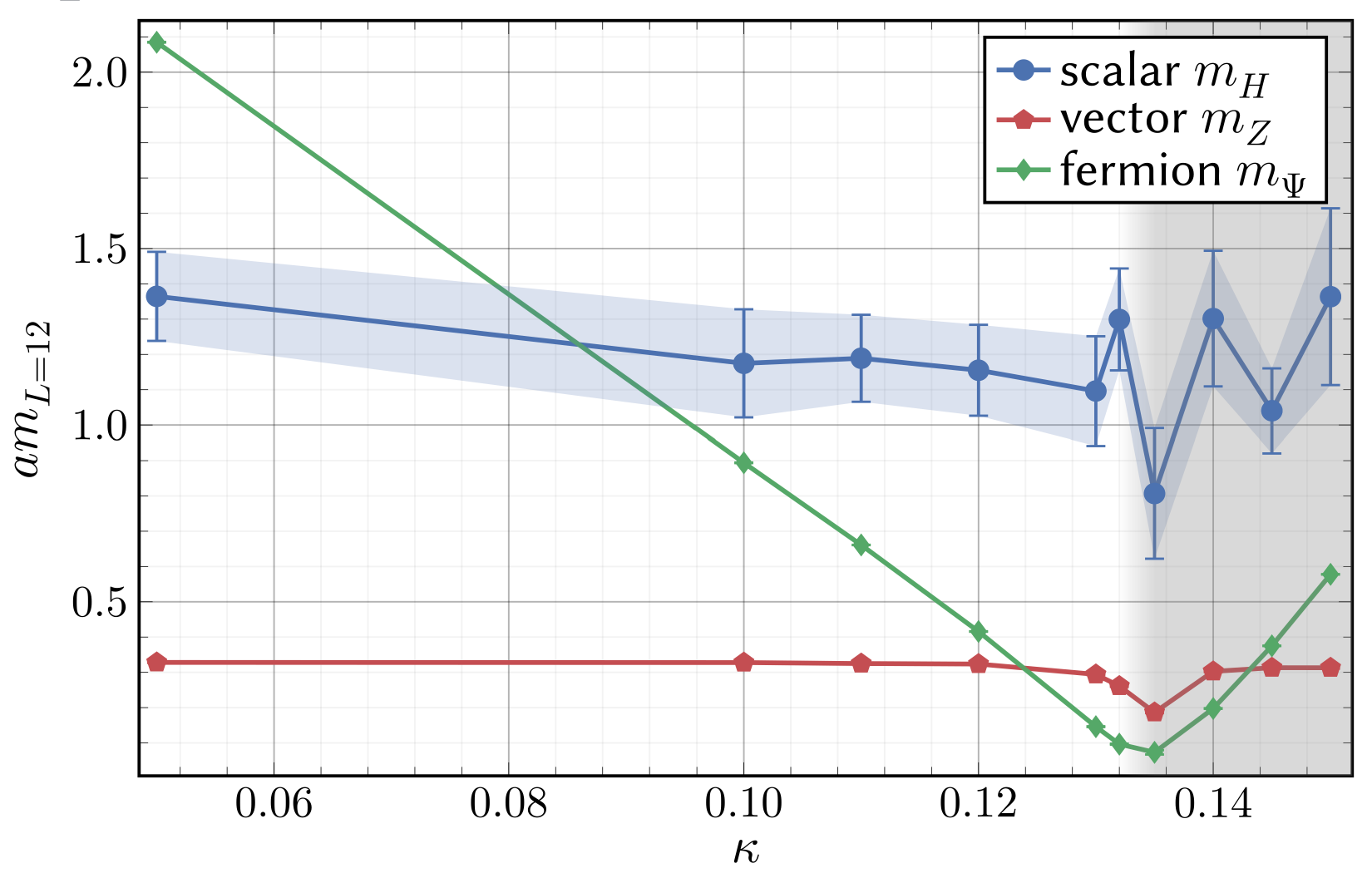}
\subcaption{Unstable Higgs mass hierarchy}
\label{fig:unstablehiggs}
\end{minipage}
\caption{Comparison of parameter spectra with a stable and unstable Higgs in the Higgs-like regime: Introducing fermions does not have a large effect on the spectrum generated by the gauge-scalar system.}
\end{figure}

To examine the phase diagram of the gauge-scalar system with added fermions, we first evaluate the dependence on the fermionic mass, which is controlled by the Wilson fermion hopping parameter $\kappa_{\mathrm{F}} = (2(m_\mathrm{F} + 4))^{-1}$. In Figs.~\ref{fig:stablehiggs} and ~\ref{fig:unstablehiggs}, it is visible that the introduction of fermions has only a mild effect on the mass hierarchy of the Higgs and the W/Z states, even once the decay thresholds into fermion pairs open. In the gray-shaded area, the correlators become unphysical. We strongly suspect that the onset of violations of reflection positivity for Wilson fermions is lowered down from $\kappa_{\mathrm{F}}=0.15$ \cite{Montvay:1994cy}, due to the presence of the scalar, to the observed value of $\kappa_{\mathrm{F}}\approx 0.14$, depending on the remaining parameters $\beta, \kappa_\mathrm{s}$, and $\lambda_\mathrm{s}$.

\begin{figure}
\centering
\includegraphics[width=\linewidth]{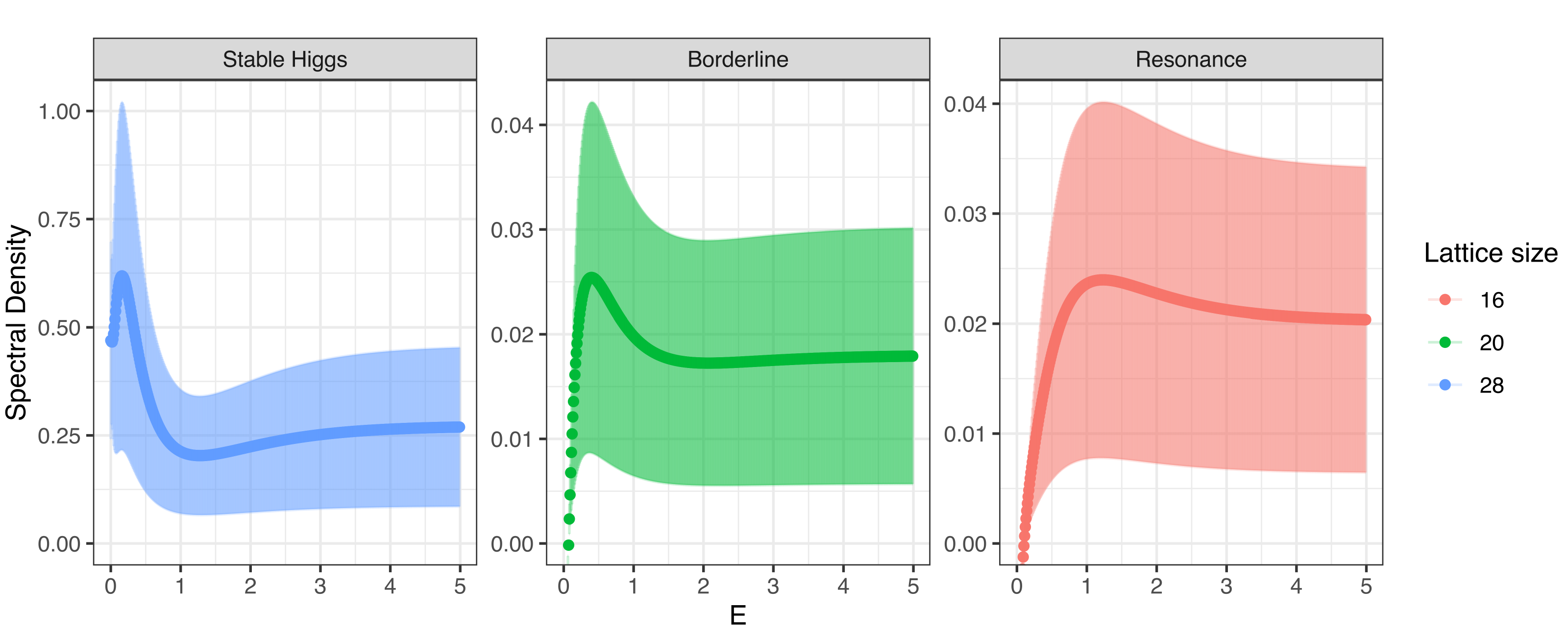}
\caption{Spectral densities extracted from the scalar operator $O_{\mathrm{Higgs}}$}
\label{fig:specdensity}
\end{figure}

The stability of the scalar can be further examined by extracting the smeared scalar spectral density function using the so-called HLT approach \cite{Hansen:2019idp}. Here, we can examine precisely the transition from a stable Higgs, which, for the smeared spectral density, gives a thin peak around the scalar mass $0.3a$, to the unstable resonance case, where elastic and inelastic threshold behavior is expected and observed (see Fig.~\ref{fig:specdensity}). In addition, a borderline ensemble shows the transition region between the two cases, where remnants of the peak are still visible, along with threshold behavior. It is a major result that the spectral density of these composite operators is compatible with the NLO APT results shown in \cite{Maas:2020kda,Dudal:2020uwb}. 

\begin{figure}
\centering
\includegraphics[width=0.5\linewidth]{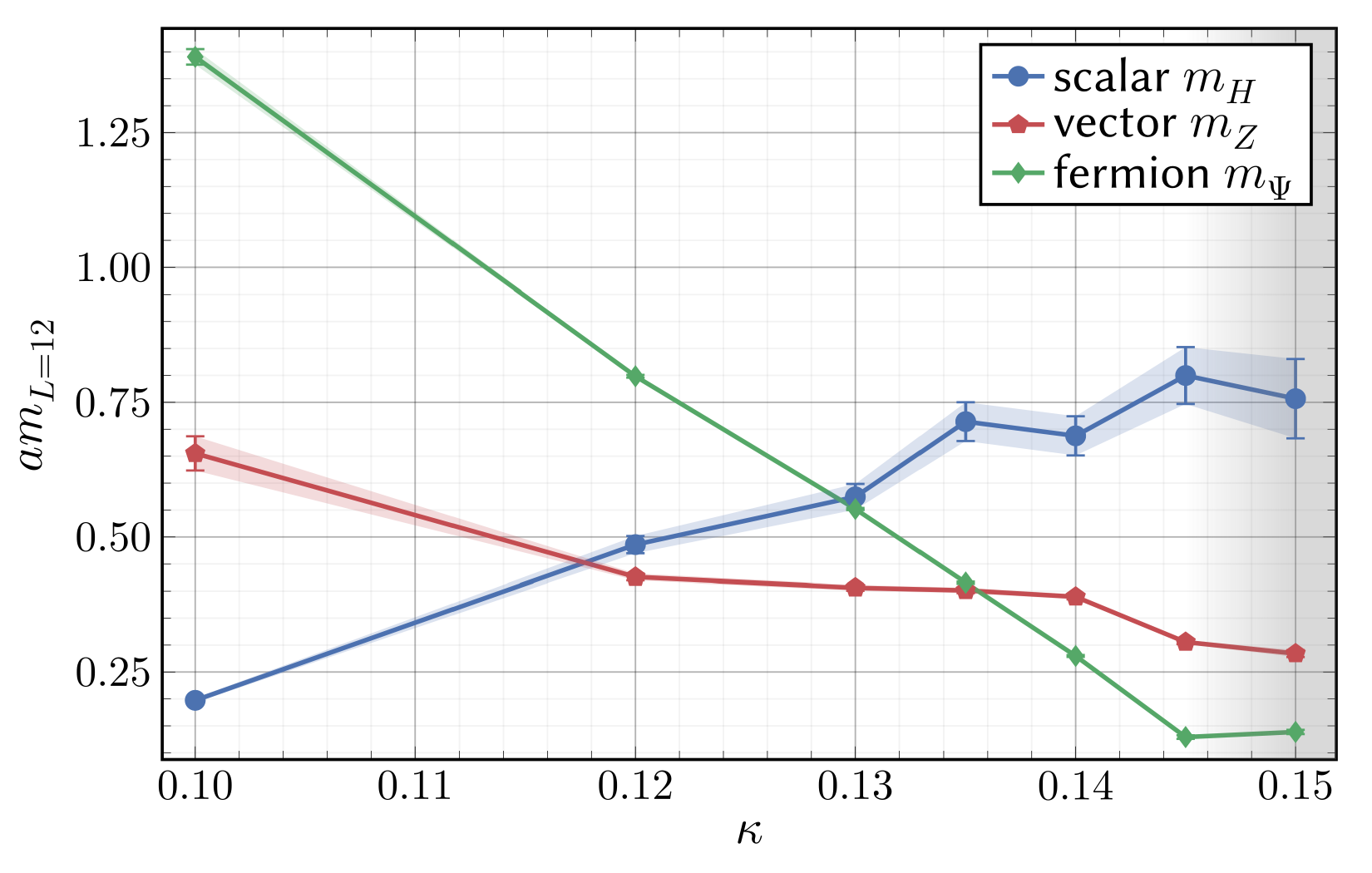}
\caption{$\kappa_F$-dependence of the mass hierarchies in the QCD-like parameter region}
\label{fig:qcdlike}
\end{figure}

A hallmark of Higgs physics on the lattice has been that whenever a Brout-Englert-Higgs effect is active, and in the absence of spontaneously broken global symmetries, the W/Z is always lighter or degenerate with the Higgs \cite{Maas:2017wzi}. No explanation for this observation is available yet. We observe, however, that for some parameters in the gauge-scalar sector, lowering the fermion mass flips the Higgs and W/Z mass hierarchy, as shown in Fig.~\ref{fig:qcdlike}. While this requires further scrutiny, there are three plausible possibilities. Either the hierarchy does not universally hold in the Brout-Englert-Higgs case, the theory moves into a QCD-like phase, or it could exhibit an Abbott-Farhi behavior \cite{Abbott:1981re}, which has also been searched for in \cite{Hansen:2017mrt} in this theory.

\section{Quasi-PDFs}\label{s:pdf}

\begin{figure}
\centering
\includegraphics[width=0.5\linewidth]{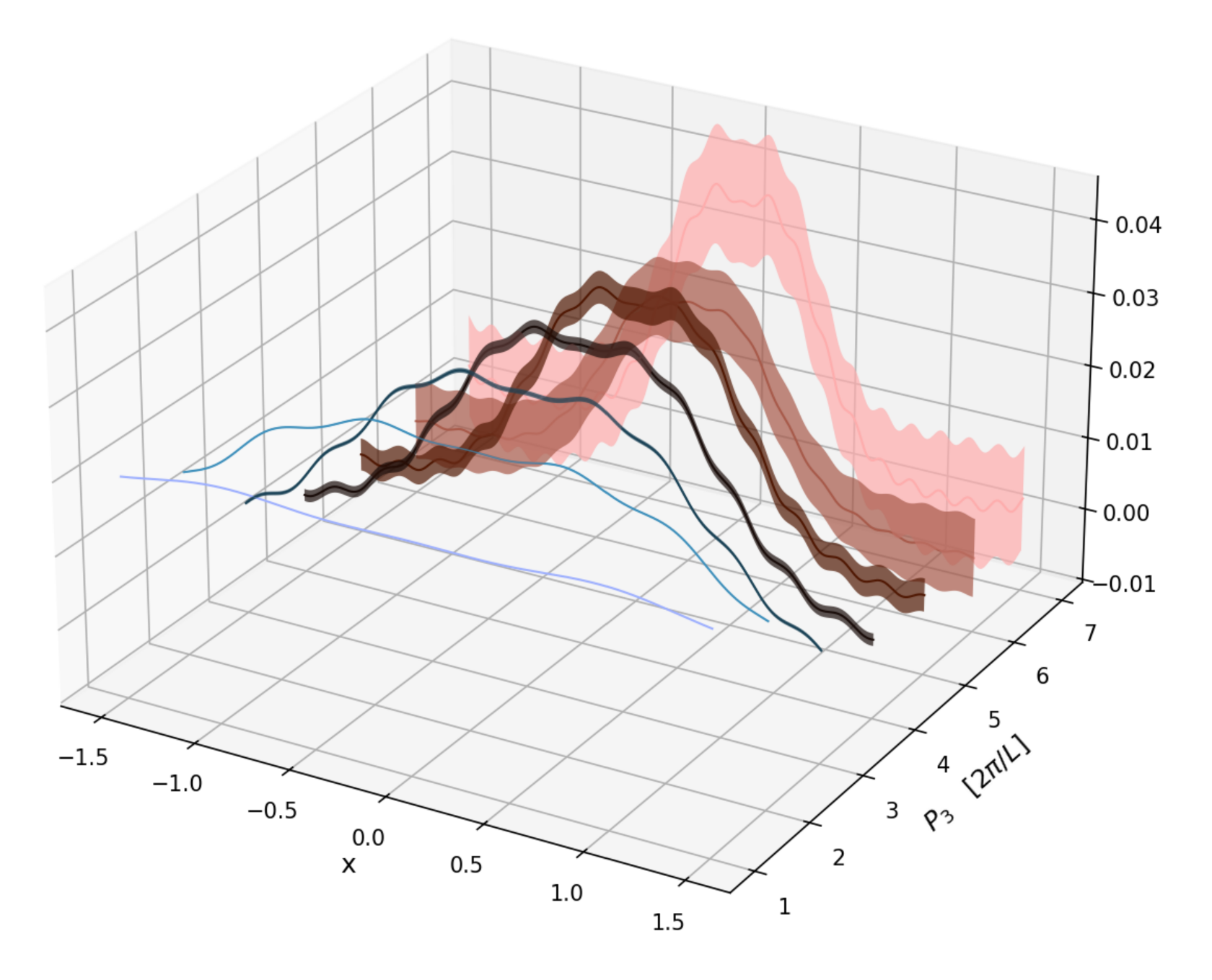}
\caption{Quasi-PDF of the W-boson with parallel polarization $x\tilde{w}^{(2)}_{W_{||}}(x, P_3)$, that shows only weak dependence on $P_3$ and a structure consistent with compositeness.}
\label{fig:wqpdf}
\end{figure}

As our asymptotic states are now composite, they have a potentially interesting internal structure. As a first step in investigating this structure, we compute the quasi-PDFs using the definition
\begin{equation}
\tilde{w}_i(x, P_3) = \int_{-\infty}^{\infty}
\dfrac{\mathrm{d}z}{2\pi xP_3}e^{-\mathrm{i}zxP_3}M_i^{W}(P_3, z) \stackrel{\tiny\mathrm{discrete}}{\sim} \dfrac{1}{2\pi x}\sum_{z=-z_\mathrm{max}}^{z_{\mathrm{max}}}e^{-\mathrm{i}zxP_3}M_{i}^{W}(P_3, z)
\end{equation}
with the matrix element obtained by fitting the ratio of the two-point and three-point correlation functions
\begin{equation}
\mathcal{R}(\bm{P};t_s) = \sum_{\tau=1}^{t_s-1}\dfrac{C^{\mathrm{3pt}}(\tau, t_s;P_3,z)}{C^{\mathrm{2pt}}(t_s;P_3)} = C + M_i^{j}(P_3,z)t_s + \mathcal{O}\left(e^{-(E_1-E_0)t_s}\right)\,.
\end{equation}
The difference between the quasi-PDF and PDF \cite{Cichy:2018mum} is suppressed by the coupling strength, which is generally much smaller in weak physics than in QCD applications. If we were examining an elementary particle, the PDF would have the shape of two Dirac delta peaks at $x=\pm 1$, but mildly smeared by the sea. There should be no significant dependence on the momentum. 

We consider the longitudinally polarized W/Z within the physical W/Z. This can be derived from the same operators as those used for gluon PDFs \cite{HadStruc:2021wmh} and the ones given in (\ref{ops}). The structure we observe in Fig.~\ref{fig:wqpdf} is obtained from a mildly noise-reduced current operator 
\begin{equation}
O^{(2)}_{\mathrm{w},||}(z) = \sum_{\mu\neq 3}\tr \left(W_{\mu3}(z)\Omega(z,0)W_{\mu3}\Omega(0,z)\right)
\end{equation}
with
\begin{equation}
W_{\mu\nu}=\frac{i}{8}\left(U_{\mu\nu}-U_{\nu\mu}+U_{\nu-\mu}-U_{-\mu\nu}+U_{-\mu-\nu}-U_{-\nu-\mu}+U_{-\nu\mu}-U_{\mu-\nu}\right)\,,
\end{equation}

\noindent where $\Omega$ is a Wilson line and $U_{\mu\nu}$ the usual plaquette. As expected, the residual momentum dependence diminishes quickly even for small momenta. The resulting (quasi-)PDF is quite different from the expected trivial structure. Thus, our results indicate an involved internal structure.

\section{Exclusive cross sections}\label{s:xs}

\begin{figure}
\centering
\includegraphics[width=0.6\linewidth]{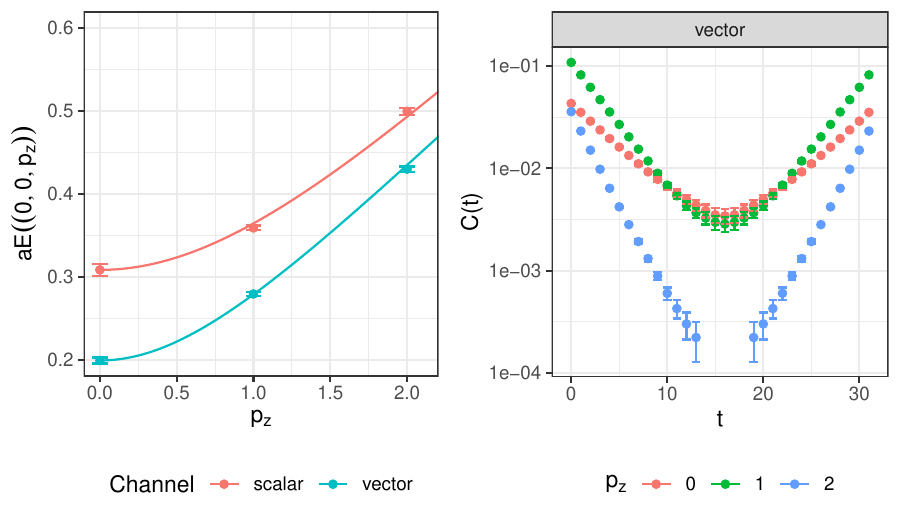}
\caption{Momentum dependence of mass and correlation functions, matched to the approximate naive behavior given by the dispersion relation}
\label{fig:momentum-dep}
\end{figure}

\begin{figure}
\centering
\includegraphics[width=0.8\linewidth]{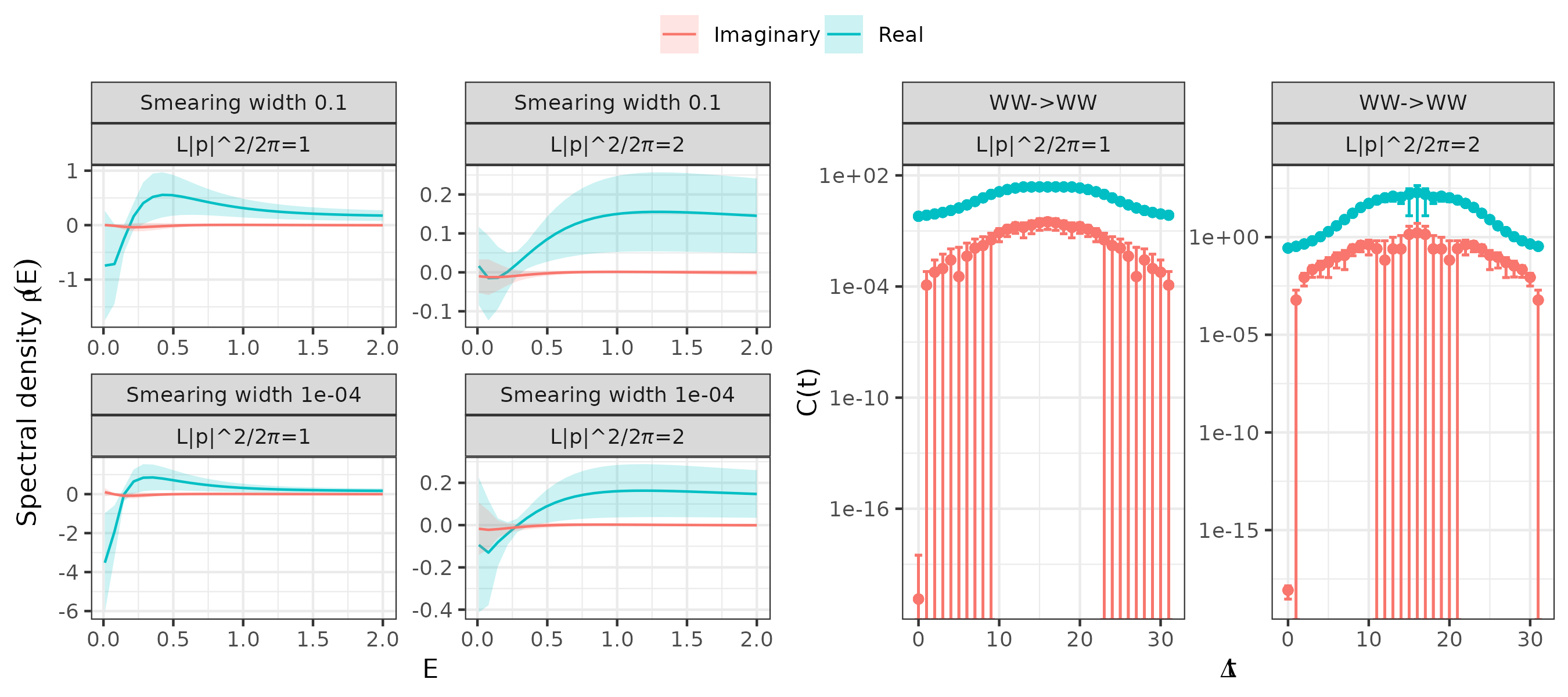}
\caption{Ratio correlation functions $C^{L=32}_{\bm{p}_4\bm{p}_1}(\bm{p}_2, \tau_2)$ and corresponding spectral densities at two different smearing widths evaluated with the HLT method using the complex kernel from \cite{Bulava:2019kbi}.}
\label{fig:ratio-and-spec-decomp}
\end{figure}

Scattering is currently most frequently examined by using the L{\"u}scher formalism \cite{Luscher:1986pf, Luscher:1990ux, Luscher:1991cf}, see also \cite{Sharpe:2026mtt} in this proceedings. This method becomes increasingly involved at the inelastic threshold, limiting its description of heavier particle and multi-particle interactions. To transcend this limitation, new methods have been developed in \cite{Bulava:2019kbi} and \cite{Patella:2024cto}, allowing for a more general computation of scattering amplitudes from the lattice directly, rather than via phase shifts. In the following, we will present results obtained using the method from \cite{Bulava:2019kbi}, but we aim to corroborate these results with \cite{Patella:2024cto} and further work below threshold with the L{\"u}scher formalism. This will allow very good systematic control. The approach of \cite{Bulava:2019kbi} depends on the evaluation of spectral densities, which were evaluated with \cite{Hansen:2019idp} and some results already shown in Fig.~\ref{fig:specdensity}.

As a first step, we are examining exclusive processes below the inelastic threshold in vector boson scattering. For the scalar channel, this was already done using the L\"uscher method in  \cite{Jenny:2022atm}, giving us the possibility to compare. Deviations from standard perturbation theory near the elastic threshold were previously observed.

To this end, we compute the boosted spectrum of the ground state in the vector channel, to obtain $E(p)$ and the LSZ factor $Z^{1/2}(p)$, see Fig.~\ref{fig:momentum-dep}. To compute the cross section at a given lattice extent $L$ and smearing width $\sigma$, we have to determine the matrix element from the spectral decomposition
\begin{equation}
\mathrm{d}\sigma(\theta, \sqrt{s}) = \dfrac{|\mathcal{M}(\bm{p}^2, \theta)|^2}{64\pi^2 \sqrt{s}^2}\,,
\end{equation}
where the matrix element is given by
\begin{equation}
\mathrm{i}\mathcal{M}_{L,\sigma} = \dfrac{E(\bm{p}_3)E(\bm{p}_2)}{Z^{1/2}(\bm{p}_3)Z^{1/2}(\bm{p}_2)}\sigma^2\widehat{\rho}^{L,\sigma}_{\bm{p}_4\bm{p}_1}(E(\bm{p}_2), \bm{p}_2) 
\end{equation}
with the smeared spectral density
\begin{equation}
\widehat{\rho}^{\sigma}_{\bm{p}_1\bm{p}_4}(q_2) = \int_{0}^{\infty}\dfrac{\mathrm{d}E_2}{\pi}\dfrac{\mathrm{i}}{E(\bm{p}_1) + q_2^0 - E_2+\mathrm{i}\sigma}\rho_{\bm{p}_1\bm{p}_4}(E_2, \bm{p}_2)
\end{equation}
computed using HLT from the ratio of a four-point and two two-point correlation functions, denoted by $C^{L}_{\bm{p}_4\bm{p}_1}(\bm{p}_2, \tau_2)$. We show the spectral densities and ratios of correlation functions in Fig.~\ref{fig:ratio-and-spec-decomp}. It is visible that it will be viable to obtain the corresponding cross sections with this level of accuracy in the simulations. These will be reported elsewhere.

\section{Conclusion and Outlook}

We reported here a number of novel results for a proxy theory of the Standard Model lepton and Higgs sector. These included preliminary results on particle-mass hierarchies depending on the lattice-ensemble parameter choice, especially the fermion hopping parameter, as well as intermediate results for computing cross sections for vector-boson scattering and PDFs.

Eventually, the program will allow us to investigate the existence of excited states in the spectrum, cross sections for scattering processes between fermions, and the substructure of composite states from their PDFs. Comparing these results to both perturbation theory and APT will yield important insights. The comparison to perturbation theory will reveal, as has already been seen in \cite{Jenny:2022atm} and conjectured on principal grounds in \cite{Maas:2022gdb}, where and how discrepancies arise. The comparison with APT will show whether these differences can be accounted for analytically, possibly supported by PDFs, or whether genuine non-perturbative methods will be necessary. This will be decisive for predicting Standard Model behavior and for the possibility of discovering genuine field-theoretical effects, such as Elitzur's theorem, at future colliders, especially the high-luminosity LHC and future lepton colliders. Such a discovery will mark a major step in testing quantum field theory beyond the purely perturbative regime, with far-reaching potential \cite{Maas:2023emb}.

Practical next steps in our analysis will be to increase the precision of the spectral densities and to extract scattering amplitudes using other methods. If we find excited states whose masses are compatible with additional (lepton) generations, especially when dynamically including Yukawa couplings, this will be very interesting for evaluating the mixing angles between the excited leptonic states. 

\acknowledgments

This project is supported by the Austrian Science Fund FWF under grant number PAT6443923. We gratefully acknowledge computational resources provided on the Austrian Scientific Computing (ASC5) infrastructure, the Graz Scientific Cluster (GSC1), and Hippo (DeiC Large Memory HPC / UCloud), managed by the eScience Center at the University of Southern Denmark. We use generative AI to support coding and language revision.

\bibliographystyle{JHEP}
\bibliography{literature}
\end{document}